\documentclass[11pt]{article}

\usepackage{amsfonts,amsmath,amssymb,amsthm}
\usepackage{xspace}

\newcommand{\<}{\langle}
\renewcommand{\>}{\rangle}

\newcommand{\be}{\begin{equation}}
\newcommand{\ee}{\end{equation}}
\def\ba#1\ea{\begin{align}#1\end{align}} 

\newcommand{\func}{\varphi}
\newcommand{\fsum}[1]{\Phi^{(#1)}}
\newcommand{\fsumt}[1]{\hat\Phi^{(#1)}}
\newcommand{\mat}{\mu}
\newcommand{\msum}[1]{M^{(#1)}}
\newcommand{\car}[2]{\chi_{#1}(#2)}
\newcommand{\carbar}[2]{\bar\chi_{#1}(#2)}

\newcommand{\Z}{{\mathbb Z}}
\newcommand{\Zp}{\Z_p}

\newcommand{\ZN}{\Z_N}

\newcommand{\N}{\mathbb{N}}
\newcommand{\C}{\mathbb{C}}
\renewcommand{\i}{\mathrm{i}}

\newcommand{\E}{\mathop{\mbox{$\mathbb{E}$}}}
\newcommand{\poly}{\mathrm{poly}}
\newcommand{\tr}{\mathop{\mathrm{tr}}\nolimits}

\newcommand{\Aut}{\mathop{\mathrm{Aut}}\nolimits}
\newcommand{\GL}{\mathop{\mathrm{GL}}\nolimits}
\newcommand{\Prsuc}{\Pr(\text{\rm success})}
\newcommand{\semidirect}{\rtimes}
\newcommand{\normalin}{\trianglelefteq}
\newcommand{\notnormalin}{\ntrianglelefteq}
\newcommand{\HSP}{\textsc{hsp}\xspace}
\newcommand{\PGM}{\textsc{pgm}\xspace}
\newcommand{\POVM}{\textsc{povm}\xspace}

\newlength{\vdotsheight}
\newcommand{\vdotsstrut}{\rule{0pt}{\vdotsheight}}

\newtheorem{theorem}{Theorem}
\newtheorem{lemma}[theorem]{Lemma}

\newcommand{\eprint}[1]{arXiv:#1}

\title{\vspace{-3ex}
       From optimal measurement to efficient quantum algorithms for
       the hidden subgroup problem over semidirect product groups}

\author{
  Dave Bacon\footnote{Santa Fe Institute, Santa Fe, NM 87501, USA}\\
  dabacon@santafe.edu
  \and
  Andrew M. Childs\footnote{Institute for Quantum Information,
  California Institute of Technology,
  Pasadena, CA 91125, USA}\\
  amchilds@caltech.edu
  \and
  Wim van Dam\footnote{Department of Computer Science,
  University of California, Santa Barbara,
  Santa Barbara, CA 93106, USA}\\
  vandam@cs.ucsb.edu
}

\date{}

\begin{document}

\maketitle


\vspace{-2ex}
\begin{abstract}
We approach the hidden subgroup problem by performing the so-called
pretty good measurement on hidden subgroup states.  For various groups
that can be expressed as the semidirect product of an abelian group
and a cyclic group, we show that the pretty good measurement is
optimal and that its probability of success and unitary implementation
are closely related to an average-case algebraic problem.  By solving
this problem, we find efficient quantum algorithms for a number of
nonabelian hidden subgroup problems, including some for which no
efficient algorithm was previously known: certain metacyclic groups as
well as all groups of the form $\Zp^r \semidirect \Zp$ for fixed $r$
(including the Heisenberg group, $r=2$).  In particular, our results
show that entangled measurements across multiple copies of hidden
subgroup states can be useful for efficiently solving the nonabelian
\HSP.
\end{abstract}

\section{Introduction}

The hidden subgroup problem (\HSP) stands as one of the major
challenges for quantum computation.  Shor's discovery of an efficient
quantum algorithm for factoring and calculating discrete logarithms
\cite{Shor:94a}, which essentially solves the abelian \HSP
\cite{Boneh:95a,Kitaev:95a}, focused attention on the question of what
computational problems might be solved asymptotically faster by
quantum computers than by classical ones.  In particular, we would
like to understand when the \emph{nonabelian} hidden subgroup problem
admits an efficient quantum algorithm.

Considerable progress on this question has been made since Shor's
discovery.  Efficient quantum algorithms have been found for the case
where the hidden subgroup is promised to be normal and there is an
efficient quantum Fourier transform over the group
\cite{Hallgren:00a}, or where the group is ``almost abelian''
\cite{Grigni:00a} or, more generally, ``near-Hamiltonian''
\cite{Gavinsky:04a}.  In addition, efficient algorithms have been
found for several groups that can be written as semidirect products of
abelian groups: the wreath product $\Z_2^n \wr \Z_2$ \cite{RB:98},
certain groups of the form $\Z_{p^k}^n \semidirect \Z_2$ for a fixed
prime power $p^k$ \cite{Friedl:03a}, $q$-hedral groups with $q$
sufficiently large \cite{Moore:04a}, and particular groups of the form
$\Z_{p^k} \semidirect \Zp$ with $p$ an odd prime \cite{IG:04}.

Unfortunately, efficient algorithms have been elusive for two cases
with particularly significant applications: the dihedral group and the
symmetric group.  An efficient algorithm for the \HSP over the
symmetric group would lead to an efficient algorithm for graph
isomorphism \cite{Boneh:95a,Ettinger:99b,Beals:97a,Hoyer:97a}, and an
efficient algorithm for the dihedral \HSP (based on the standard
approach described in Section~\ref{sec:hsp}) would lead to an
efficient algorithm for certain lattice problems \cite{Regev:02a}.
While no polynomial-time quantum algorithms for these problems are
known, Kuperberg recently gave a subexponential (but superpolynomial)
time and space algorithm for the dihedral \HSP \cite{Kuperberg:03a},
and Regev improved the space requirement to be only polynomial
\cite{Regev:04a}.

Recently, we have advocated an approach to solving hidden subgroup
problems based on a state estimation technique known as the {\em
pretty good measurement} (\PGM).  The standard approach to solving the
\HSP with a quantum computer (described in detail in
Section~\ref{sec:hsp}) reduces the problem to the quantum mechanical
task of distinguishing the members of an ensemble of \emph{hidden
subgroup states}.  For the dihedral \HSP, we showed that the \PGM is
in fact the optimal measurement for distinguishing any number of
copies of the hidden subgroup states, that it successfully identifies
the hidden subgroup with a polynomial number of copies of the states
(for which we gave a tight lower bound), and that its implementation
is closely related to an average-case subset sum problem
\cite{BCD:05a}.  Unfortunately, this subset sum problem appears to be
difficult, so the approach did not yield an efficient algorithm for
the dihedral \HSP.

In this paper, we continue our study of the \PGM as a tool for solving
hidden subgroup problems.  We apply the method to all groups that can
be written as a semidirect product $A \semidirect \Zp$ of an abelian
group $A$ and a cyclic group of prime order $p$.  As in the case of
the dihedral group, the \PGM for these groups is closely related to a
certain kind of average-case algebraic problem, which we call the {\em
matrix sum problem}.  For some groups, the matrix sum problem can be
solved efficiently, leading to an efficient algorithm for the \HSP.
We demonstrate this for two classes of groups in
Sections~\ref{sec:metacyclic} and \ref{sec:heisesque}.  In
Section~\ref{sec:metacyclic}, we give an efficient algorithm for
metacyclic groups $\ZN \semidirect \Zp$ with $N/p = \poly(\log N)$,
generalizing the result of \cite{Moore:04a} for $p$-hedral groups,
which requires $N$ prime.  In Section~\ref{sec:heisesque}, we give an
efficient algorithm for any group of the form $\Zp^r \semidirect \Zp$
for fixed $r$, including the Heisenberg group, the unique nontrivial
semidirect product $\Zp^2 \semidirect \Zp$.  For the groups $\Zp^r
\semidirect \Zp$, the matrix sum problem is a system of polynomial
equations over a finite field, which can be solved efficiently using
Buchberger's (classical) algorithm for computing Gr\"obner bases
\cite{CLO:97}.  For the metacyclic groups, the matrix sum problem
requires the calculation of discrete logarithms, which can be done
efficiently using Shor's algorithm \cite{Shor:94a}.  In both cases,
the algorithm uses abelian Fourier transforms, but does not explicitly
use a nonabelian Fourier transform.

To simplify the \PGM approach to the \HSP, we find it useful to focus
on a specific set of subgroups rather than allowing the hidden
subgroup to be arbitrary.  For the dihedral group, Ettinger and
H{\o}yer showed that it is sufficient to consider the case where the
hidden subgroup is either trivial or has order $2$.  In
Section~\ref{sec:reduction}, we give an analogous reduction showing
that it is sufficient to consider subgroups that are either trivial or
cyclic and of order $p$.  In fact, this reduction alone is sufficient
to solve the \HSP over some nonabelian groups, such as $\Z_2^n \wr
\Z_2$ (cf.\ \cite{RB:98}) and the groups $P_{p,r}$ of \cite{IG:04}.

It is well known that the query complexity of the hidden subgroup
problem is polynomial, and in particular, that only polynomially many
copies of the hidden subgroup states are sufficient to solve the
problem in general \cite{Ettinger:04a}.  However, measurements that
operate on a single hidden subgroup state at a time are in general
{\em not} sufficient---in particular, they are insufficient for the
symmetric group \cite{Moore:05a}.  Kuperberg's algorithm for the for
the dihedral \HSP depends essentially on using entangled measurements
on multiple copies of the hidden subgroup states \cite{Kuperberg:03a},
but unfortunately does not run in polynomial time.  As far as we know,
our algorithm for the \HSP over groups of the form $\Zp^r \semidirect
\Zp$ is the first efficient quantum algorithm to use entangled
measurements across multiple copies (specifically, $r$ copies) of the
hidden subgroup states.  This result provides hope that the
distinguishability of polynomially many copies of hidden subgroup
states may lead to further efficient algorithms through the
implementation of entangled measurements.

The remainder of the article is organized as follows.  In
Section~\ref{sec:hsp} we review the hidden subgroup problem and the
standard approach to solving it with a quantum computer, and in
Section~\ref{sec:semidirect} we review some relevant facts about
semidirect product groups.  Then, in Section~\ref{sec:reduction}, we
give the reduction to cyclic subgroups.  In Section~\ref{sec:pgm}, we
present the pretty good measurement approach in detail, describing the
relationship between hidden subgroup states and the matrix sum
problem, computing the success probability of the \PGM, proving its
optimality, and explaining how to implement it on a quantum computer.
We apply the approach to specific groups in
Sections~\ref{sec:metacyclic} and \ref{sec:heisesque}, and we conclude
in Section~\ref{sec:discussion} with a discussion of the results and
some open problems.

\section{Definitions}

\subsection{Hidden subgroup problem}
\label{sec:hsp}

Let $G$ be a finite group.  We say that a function $f:G \to S$ (where
$S$ is a finite set) hides the subgroup $H \le G$ if $f$ is constant
and distinct on left cosets of $H$ in $G$.  The hidden subgroup
problem is the following: given the ability to query the function $f$,
find a generating set for $H$.

To approach the \HSP with a quantum computer, we must have the ability
to query $f$ in superposition.  In particular, we are provided with a
quantum oracle $U_f$ acting as $  U_f: |g,y\> \mapsto |g,y \oplus
f(g)\>$ for all $g \in G$ and $y \in S$, where $\oplus$ denotes the
bitwise exclusive or operation, and the elements of $S$ are
represented by bit strings of length $\poly(\log|G|)$.  An efficient
quantum algorithm for the \HSP is an algorithm using this black box
that finds a generating set for $H$ in time $\poly(\log|G|)$.

The standard approach to solving the \HSP on a quantum computer is as
follows.  Create a superposition over all elements of the group and
then query the function in superposition, giving
\be
  U_f: \frac{1}{\sqrt{|G|}} \sum_{g \in G} |g,0\>
  \mapsto
  \frac{1}{\sqrt{|G|}} \sum_{g \in G} |g,f(g)\>
\,.
\ee
Next, discard the second register, leaving the first register in a
mixed state whose form depends on the hidden subgroup $H$,
\be
  \rho_H := \frac{|H|}{|G|} \sum_{g \in K} |gH\>\<gH|
\quad \text{with the coset states}\quad
  |gH\> := \frac{1}{\sqrt{|H|}} \sum_{h \in H} |gh\>\,,
\label{eq:hss}
\ee
where $K \subset G$ is a complete set of left coset representatives of
$H$ in $G$.  We call $\rho_H$ the \emph{hidden subgroup state} hiding
the subgroup $H$.  In the standard approach, one attempts to determine
$H$ using samples of hidden subgroup state $\rho_H$.

\subsection{Semidirect product groups}
\label{sec:semidirect}

The semidirect product of two groups $A,B$ is defined in terms of a
homomorphism $\varphi:B \to \Aut(A)$, where $\Aut(A)$ denotes the
automorphism group of $A$.  The semidirect product group $A
\semidirect_\varphi B$ consists of the elements $(a,b)$ with $a \in A$
and $b \in B$.  With group operations in $A$ and $B$ written
additively, the group operation in $A \semidirect_\varphi B$ is
defined as $(a,b)(a',b') = (a+\varphi(b)(a'),b+b')$.  It is not hard
to show that group inversion satisfies $(a,b)^{-1} =
(\varphi(-b)(-a),-b)$.

We consider the hidden subgroup problem for semidirect product groups
$G=A \semidirect_\varphi \Zp$, where $A$ is an abelian group and $p$
is prime.  In this case, since $\varphi$ is a homomorphism of a cyclic
group, it is determined entirely by $\varphi(1)$ (in particular,
$\varphi(p)$ is the identity map); hence, with a slight abuse of
notation, we let $\func:=\varphi(1)$ (an automorphism of $A$)
henceforth, and define $\func^b := ({\varphi \circ \cdots \circ
\varphi})$ ($b$ times), giving $\varphi(b) = \func^b$.

We will be especially interested in cyclic subgroups of such a
semidirect product group.  For any $a \in A$, we have
$ \<(a,1)\> =
  \{(0,0),(a,1),(a+\func(a),2),(a+\func(a)+\func^2(a),3),\ldots\}$.
For convenience, we introduce the function $\fsum{b}: A \to A$ defined
by
\be
  \fsum{b}(a) := \sum_{i=0}^{b-1} \func^i(a)
\ee
such that the elements of $\<(a,1)\>$ are $(a,1)^b = (\fsum{b}(a),b
\bmod p)$ for $b \in \N$.

\section{Reduction to cyclic subgroups}
\label{sec:reduction}

To simplify the hidden subgroup problem over semidirect product
groups, we can reduce the problem to that of finding either the
trivial hidden subgroup or a hidden cyclic subgroup of order $p$.
This reduction generalizes a result of Ettinger and H{\o}yer, who
reduced the dihedral \HSP to the problem of finding a trivial hidden
subgroup or a hidden reflection (an order $2$ subgroup).

\begin{lemma}[cf.\ Theorem~2.3 of \cite{Ettinger:00a}]
\label{lem:reduction}
To find an efficient algorithm for the \HSP over $A \semidirect \Zp$
with $p$ prime, it suffices to find an efficient algorithm for the
\HSP over $A_2 \semidirect \Zp$ for any $A_2 \le A$ with the promise
that either $H$ is trivial or $H=\<(d,1)\>$ for some $d \in A_2$ with
$|H|=p$.
\end{lemma}
\begin{proof}
The proof proceeds along the lines of the Ettinger-H{\o}yer reduction.
Their idea is to factor out the part of the hidden subgroup that lies
within $A$.  For generalized dihedral groups ($p=2$, $\func(a)=-a$),
their reduction goes through unchanged, as noted in \cite{Friedl:03a}.
However, in some cases there is an additional complication coming from
the fact that the $A$-part of the hidden subgroup may not be normal in
$G$.  Fortunately, we can deal with this case separately.

Let $f: G \to S$ hide a subgroup $H$.  Let $G_1 := A \times \{0\}$,
and let $H_1 := H \cap G_1 =: A_1 \times \{0\}$.  Since $f$ restricted
to the abelian group $G_1$ hides the subgroup $H_1$, we can
efficiently find generators for $H_1$ by solving an abelian hidden
subgroup problem.

We would like to factor out $H_1$, so we check whether it is normal in
$G$.  For $g=(a,b) \in G$ and $h_1=(h,0) \in H_1$, we have
$ g h_1 g^{-1} = (a,b)(h,0)(\func^{-b}(-a),-b)
               = (\func^b(h),0)$.
Therefore $H_1 \normalin G$ if and only if $\func(H_1) = H_1$, where
$\func(H_1) := \{(\func(h),0): h \in A_1\}$.  If $H_1 \ne H$, then
$\func(H_1) = H_1$: since $p$ is prime, there must be some $(d,1) \in
H$, and for any $h \in A_1$, $(d,1)(h,0)(d,1)^{-1} = (\func(h),0) \in
H$, hence $\func(h) \in A_1$.

If $H_1 = H$, it could be that $\func(H_1) \ne H_1$.  However, given a
generating set for $H_1$, we can check whether $H_1 \normalin G$ (for
example, we can use the results of \cite{Watrous:01}; note that $G$ is
solvable since its commutator subgroup is abelian\footnote{In some
cases, the results of \cite{Watrous:01} may not be required; for
example, in the case $A = \Zp^r$ considered in
Section~\ref{sec:heisesque}, $H_1$ can be viewed as a subspace, and it
is sufficient to check whether this subspace is invariant under the
action of the matrix $\mat$ that defines the automorphism $\func$.}).
If we find $H_1 \notnormalin G$, then we know that $H=H_1$, and we are
done.  Otherwise, we learn that $H_1 \normalin G$, and we proceed to
factor out $H_1$.

The (left) cosets of $H_1$ in $G$ can be represented as follows: for
$g=(a,b) \in G$,
\be
  g H_1 = \{(a,b)(h,0): h \in A_1\}
        = \{(a+\func^b(h),b): h \in A_1\}
        = \{(a+h,b): h \in A_1\}
\,.
\ee
Thus the cosets can be labeled by $b \in \Zp$ and $a \in A_2 :=
A/A_1$.  Now we work in the group $G_2 := G/H_1 \cong A_2
\semidirect_{\varphi_2} \Zp$ whose elements are these cosets.  Note
that $G_2$ inherits its defining automorphism $\varphi_2$ from the
original automorphism $\varphi$.

Since $f$ is constant on cosets of $H_1 \le H$, we can consider it as
a function $f: G_2 \to S$ with the hidden subgroup $H_2 := H/H_1$.  If
$H_1=H$ then $H_2$ is trivial.  If $H_1 \ne H$, then we must have $H =
\<H_1, (a,1)\>$ for some $a \in A$, which can be seen as follows.
Since $p$ is prime, there must be some $(a,1) \in H$; but for any
additional $(a',1) \in H$, we have $(a',1)(a,1)^{-1} = (a'-a,0) \in
H_1$, and hence $(a',1) \in \<H_1,(a,1)\>$. Also, note that $\<(a,1)\>
\cap H_1 = \<(\fsum{p}(a),0)\>$. Thus, by the second isomorphism
theorem, $H_2 = H/H_1 \cong \<(a,1)\> / \<(\fsum{p}(a),0)\>$, which is
a cyclic group of order $p$ generated by $(d,1)$ for some $d \in A_2$.
\end{proof}

From now on we assume that the hidden subgroup is $H = \<(d,1)\> =
\{(\fsum{b}(d),b): b \in \Zp\}$ for some $d \in A$.  Note in
particular that since $|H|=p$, we have $\fsum{p}(d)=0$.

This reduction alone is enough to solve the \HSP over some semidirect
product groups.  For example, the \HSP over the wreath product groups
$\Z_2^n \wr \Z_2$ (for which an efficient quantum algorithm was given
in \cite{RB:98}) and $\Z_{p^r} \semidirect_\varphi \Z_p$ with
$\varphi(a)=(p^{r-1}+1)a$ (for which an efficient quantum algorithm
was given in \cite{IG:04}) are both reduced to instances of the
abelian \HSP.  However, in general, we are left with a nonabelian
\HSP, which we attempt to solve using the pretty good measurement.

\section{The pretty good measurement approach}
\label{sec:pgm}

In this section, we present the pretty good measurement approach to
the hidden subgroup problem over $G = A \semidirect \Zp$.  We begin in
Section~\ref{sec:hsstates} by describing the hidden subgroup states
and expressing them in terms of a certain algebraic problem called the
matrix sum problem.  Then, in Section~\ref{sec:meas}, we describe the
pretty good measurement for distinguishing these states.  In
Section~\ref{sec:sucprob}, we give an expression for the success
probability of the measurement (as well as general upper and lower
bounds), and in Section~\ref{sec:opt}, we prove that the measurement
is optimal.  Finally, in Section~\ref{sec:implement}, we explain how
the measurement can be implemented by solving the matrix sum problem.

\subsection{Hidden subgroup states and the matrix sum problem}
\label{sec:hsstates}

According to (\ref{eq:hss}), the hidden subgroup states are uniform
mixtures of uniform superpositions over the left cosets of $H$ in $G$.
A complete set of left coset representatives of $H=\<(d,1)\>$ in $G =
A \semidirect \Zp$ is given by $L = \{(\ell,0): \ell \in A\}$, and we
have the coset states
\be 
  |\psi_{\ell,d}\> := \frac{1}{\sqrt p} \sum_{b \in \Zp}
                    |(\ell,0)(\fsum{b}(d),b)\>
                =  \frac{1}{\sqrt p} \sum_{b \in \Zp}
                    |\ell + \fsum{b}(d),b\>
\,.
\label{eq:cosetstates}
\ee
The hidden subgroup state is
\be
  \rho_d = \frac{1}{|A|} \sum_{\ell \in A} |\psi_{\ell,d}\>\<\psi_{\ell,d}|
\,.
\ee
Fourier transforming the first register (over $A$) gives
\be
  \tilde\rho_d = \frac{1}{|A|} \sum_{x \in A}
                  |\tilde\psi_{x,d}\>\<\tilde\psi_{x,d}|
\quad \text{where} \quad
  |\tilde\psi_{x,d}\> := \frac{1}{\sqrt{p}} \sum_{b \in \Zp}
                         \car{x}{\fsum{b}(d)} |x,b\>
\label{eq:fouriercoset}
\ee
and where $\chi_x: A \to \C$ for $x \in A$ denotes the $x$th group
character of $A$, satisfying $\chi_x \cdot \chi_{x'} = \chi_{x+x'}$
and $\chi_x(y) = \chi_y(x)$.  (For example, for $A=\ZN$, $\car{x}{y} =
\exp(2 \pi \i xy/N)$; for $A=\Zp^r$, $\car{x}{y} = \exp(2 \pi \i (x
\cdot y)/p)$.)

By Lemma~\ref{lemma:conjPhi} in Appendix~\ref{app:matrep}, there
exists a function $\fsumt{b}:A\rightarrow A$ such that
$\car{x}{\fsum{b}(d)} = \car{\fsumt{b}(x)}{d}$ for all $d,x\in A$;
hence
\be
  \tilde\rho_d = \frac{1}{|G|} \sum_{x \in A} \sum_{b,c \in \Zp}
           \car{\fsumt{b}(x)}{d} \, \carbar{\fsumt{c}(x)}{d} \,
           |x,b\>\<x,c|
\,.
\label{eq:hsstate}
\ee
For $k$ copies, we have simply
\be
 \tilde\rho_d^{\otimes k}
    = \frac{1}{|G|^k} \sum_{x \in A^k} \sum_{b,c \in \Zp^k}
      \car{\fsumt{b}(x)}{d} \, \carbar{\fsumt{c}(x)}{d} \,
      |x,b\>\<x,c|
\ee
where for $b \in \Zp^k$ and $x \in A^k$,
\be
  \fsumt{b}(x) := \sum_{j=1}^k \fsumt{b_j}(x_j)
\,.
\ee
To simplify this expression further, for any $x \in A^k$ and $w \in
A$, let $S^x_w := \{b \in \Zp^k: \fsumt{b}(x) = w\}$ denote the set of
solutions to the equation $\fsumt{b}(x)=w$, let $\eta^x_w := |S^x_w|$
denote the number of such solutions, and let
\be
  |S^x_w\> := \frac{1}{\sqrt{\eta^x_w}} \sum_{b \in S^x_w} |b\>
\ee
denote the (normalized) uniform superposition over all solutions.  (If
$\eta^x_w=0$, then no such state can be defined, and we use the
convention $|S^x_w\>=0$.) Then the hidden subgroup state can be
written
\be
  \tilde\rho_d^{\otimes k}
    = \frac{1}{|G|^k} \sum_{x \in A^k} \sum_{w,v \in A}
      \car{w}{d} \, \carbar{v}{d} \,
      \sqrt{\eta^x_w \eta^x_v} \, |x,S^x_w\>\<x,S^x_v|
\label{eq:hsstates}
\,.
\ee

Clearly, the hidden subgroup states are closely connected to the
problem of finding solutions $b \in \Zp^k$ to the equation
\be
  \fsumt{b}(x)=w
\,,
\ee
where $x\in A^k$ and $w \in A$ are chosen uniformly at random.  We
refer to this problem as the \emph{matrix sum problem} because we can
represent $\func$ as a matrix $\mat$, and hence $\fsumt{b}$ as a sum
of matrices $\msum{b}$ (see Appendix~\ref{app:matrep}).  In the case
$A=\ZN$ of Section~\ref{sec:metacyclic}, the matrix sum is a single
scalar $\msum{b} \in \ZN^\times$, while in the case $A=\Zp^r$ of
Section~\ref{sec:heisesque}, $\msum{b} \in \Zp^{r \times r}$ is the
transpose of the matrix representing $\fsum{b}$.

In general, the matrix sum problem will have many solutions when $k$
is large, and few solutions when $k$ is small.  Since $\sum_{w \in A}
\eta^x_w = p^k$, the expected number of solutions is
\be
  \E_{{x \in A^k, w \in A}}[\eta^x_w] = \frac{p^k}{|A|}
\,.
\label{eq:exeta}
\ee
Thus, we typically expect the matrix sum problem to have many
solutions for $k \gg \log_p |A|$ and few solutions for $k \ll \log_p
|A|$.  In fact, we often find a sharp transition at $k \sim \log_p
|A|$.

Note that for the dihedral group of order $2N$, $A=\ZN$ and $p=2$ with
$\func(x)=-x$, the matrix sum problem is the average-case subset sum
problem \cite{BCD:05a}, while for the abelian case with $\func(x)=x$
(so that $\fsum{b}(x)=bx$), the matrix sum problem is simply $bx=w$.

\subsection{The measurement}
\label{sec:meas}

The \emph{pretty good measurement} (\PGM, also known as the
\emph{square root measurement} or \emph{least squares measurement}) is
a positive operator valued measure (\POVM) that often does a pretty
good job of distinguishing members of an ensemble of quantum states
\cite{HW:94}.  For the ensemble of states $\sigma_j$ with equal a
priori probabilities, the \PGM $\{E_j\}$ is given by
\be
  E_j := \Sigma^{-1/2} \sigma_j \Sigma^{-1/2}
\quad
\text{where}
\quad
  \Sigma := \sum_j \sigma_j
\,,
\label{eq:pgm}
\ee
and the inverse is taken over the support of $\Sigma$.  Clearly, the
\PGM is a \POVM over the support of the ensemble.

For our ensemble of hidden subgroup states, using $k$ copies of the
state, we have
\be
  \Sigma
    := \sum_{j \in A} \rho_j^{\otimes k}
    = \frac{|A|}{|G|^k} \sum_{x \in A^k} \sum_{w \in A}
       \eta^x_w \, |x,S^x_w\>\<x,S^x_w|
\label{eq:sigma}
\ee
where we have assumed for simplicity that the subgroup $\<(j,1)\>$ has
order $p$ for every $j \in A$.  Since $\Sigma$ is diagonal, the square
root of its inverse (over its support) is particularly easy to
calculate.  Inserting (\ref{eq:hsstates}) and (\ref{eq:sigma}) into
(\ref{eq:pgm}), we find the measurement operators
\be
  E_j = \frac{1}{|A|} \sum_{x \in A^k} \sum_{w,v \in A}
        \car{w}{j} \, \carbar{v}{j} \, |x,S^x_w\>\<x,S^x_v|
\,.
\label{eq:hsspgm}
\ee
This defines the pretty good measurement for distinguishing order $p$
hidden subgroup states of $G$.

To solve the \HSP using Lemma~\ref{lem:reduction}, we must also
identify the trivial subgroup.  However, if the \PGM correctly
identifies an order $p$ subgroup when one exists, we can simply look
for an order $p$ subgroup and check whether the function is constant
on the identity and some generator, which it will not be if the hidden
subgroup is in fact trivial.  Therefore, from now on we focus on
finding an order $p$ hidden subgroup.

In general, $G$ will have some subgroups $\<(j,1)\>$ of order $p$ and
some such subgroups whose orders are larger integer multiples of $p$.
In this case, (\ref{eq:hsspgm}) is not, strictly speaking, the \PGM
for distinguishing the order $p$ hidden subgroups alone.  Furthermore,
the state (\ref{eq:hsstates}) for $d$ corresponding to a non-order $p$
subgroup is not even a hidden subgroup state.  However, since
(\ref{eq:hsstates}) is always a valid quantum state, the resulting
\PGM is well defined.  It is convenient to work with this \PGM even
when not all subgroups $\<(j,1)\>$ have order $p$.  If the measurement
identifies the order $p$ hidden subgroups with reasonable probability,
then by Lemma~\ref{lem:reduction}, this is sufficient to solve the
\HSP.

\subsection{Success probability}
\label{sec:sucprob}

The probability of successfully identifying an order $p$ hidden
subgroup $\<(d,1)\>$ is independent of $d$, and is given by
\be
  \Prsuc
    := \tr E_d \rho_d^{\otimes k} 
    = \frac{p}{|G|^{k+1}} \sum_{x \in A^k}
       \Big(\sum_{w \in A} \sqrt{\eta^x_w} \Big)^2
\label{eq:sucprob}
\,.
\ee
For the \PGM to successfully solve the \HSP, this probability must not
be too small.  Whether the success probability is appreciable
essentially depends on whether the corresponding matrix sum problem
has many solutions.  Specifically, we have the following:

\begin{lemma}[Cf.\ Theorem~2 of \cite{BCD:05a}]
\label{lem:sucprob}
If $\Pr(\eta^x_w \ge \alpha) \ge \beta$ for uniformly random $x \in
A^k$ and $w \in A$ (i.e., if most instances of the matrix sum problem
have many solutions), then
$\alpha \beta^2 |A|/p^k \le \Prsuc \le p^k/|A|$.
\end{lemma}

\begin{proof}
For the upper bound, we have
\be
  \Prsuc \le \frac{p}{|G|^{k+1}} \sum_{x \in A^k}
              \Big( \sum_{w \in A} \eta^x_w \Big)^2
         =   \frac{p^k}{|A|}
\ee
since the $\eta$'s are integers and $\sum_{w \in A} \eta^x_w = p^k$
for any $x$.  For the lower bound, we have
\be
  \Prsuc \ge \frac{|A|}{p^k} \bigg( \frac{1}{|A|^{k+1}} \sum_{x \in
             A^k} \sum_{w \in A} \sqrt{\eta^x_w} \bigg)^2
\ee
by Cauchy's inequality applied to (\ref{eq:sucprob}). Now
\be
  \frac{1}{|A|^{k+1}} \sum_{x \in A^k} \sum_{w \in A} \sqrt{\eta^x_w}
  \ge \sqrt\alpha \Pr(\eta^x_w \ge \alpha) \,,
\ee
so by the hypothesis,
$\Prsuc \ge \alpha \beta^2 |A|/p^k$
as claimed.
\end{proof}

\subsection{Optimality}
\label{sec:opt}

In fact, the pretty good measurement is the optimal measurement for
distinguishing the states (\ref{eq:hsstates}), in the sense that it
maximizes the success probability (\ref{eq:sucprob}).  This can be
seen using the following result of Holevo and Yuen, Kennedy, and Lax:

\begin{theorem}[\cite{Holevo:73b,Yuen:75}]
  Given an ensemble of states $\sigma_j$ with equal a priori
  probabilities, the measurement with \POVM elements $E_j$ maximizes
  the probability of successfully identifying the state if and only if
  $\sum_i \sigma_i E_i = \sum_i E_i \sigma_i$ and $\sum_i \sigma_i E_i
  \ge \sigma_j$ for all $j$.
  \label{thm:holevo}
\end{theorem}

The optimality of the \PGM (\ref{eq:hsspgm}) can be proved by directly
verifying these conditions.  Note that this optimality does not
necessarily follow from \cite{MR:05}, as the hidden subgroups may not
be conjugates.  However, the same basic principle is at work.

Identifying the optimal measurement can be useful for proving lower
bounds on the number of hidden subgroup states required to solve the
\HSP \cite{BCD:05a}.  Of course, for the purpose of finding an
efficient algorithm, it is not necessary for the measurement to be
optimal; rather, it is sufficient for it to identify the hidden
subgroup with reasonable probability.  In fact, it may be that a
suboptimal measurement which nevertheless is sufficient to solve the
\HSP is significantly easier to implement than the optimal one.
However, it is encouraging to find that a particularly straightforward
measurement is in fact optimal, and we will see that this measurement
does lead to efficient algorithms in some cases.  Together with the
result of Ip showing that Shor's algorithm implements the optimal
measurement for the abelian hidden subgroup problem \cite{Ip:03a},
this suggests that identifying optimal measurements may be a useful
guiding principle for discovering quantum algorithms.

\subsection{Implementation}
\label{sec:implement}

To find an efficient quantum algorithm based on the \PGM, we must show
how to implement the measurement efficiently on a universal quantum
computer.  Unsurprisingly, this implementation is also closely related
to the matrix sum problem.

According to Neumark's theorem \cite{Neumark:43}, any \POVM can be
implemented by a unitary transformation on the system together with an
ancilla, followed by a measurement in the standard basis.  In
particular, for a \POVM consisting of $N$ rank one operators $E_j =
|e_j\>\<e_j|$ in a $D$-dimensional Hilbert space, $U$ has the block
form
\be
  U = \begin{pmatrix}V & X \\ Y & Z\end{pmatrix}
\ee
where the columns of the $N \times D$ matrix $V$ are the $D$-vectors
$|e_j\>$, i.e.,
$V = \sum_{j=1}^N |j\>\<e_j|$.

Recall from (\ref{eq:hsspgm}) that the \POVM operators for the \PGM on
hidden subgroup states can be written
\be
  E_j = \sum_{x \in A^k} |x\>\<x| \otimes E^x_j
\quad
\text{where}
\quad
E^x_j := |e^x_j\>\<e^x_j|
\quad
\text{with}
\quad
  |e^x_j\> := \frac{1}{\sqrt{|A|}} \sum_{w \in A} \car{w}{j} |S^x_w\>
\,.
\ee
In other words, each $E_j$ is block diagonal, with blocks labeled by
$x \in A^k$, and where each block is rank one.  Thus, the measurement
can be implemented in a straightforward way by first measuring the
block label $x$ and then performing the \POVM $\{E^x_j\}_{j \in A}$
conditional on the first measurement result.

To implement the \POVM $\{E^x_j\}_{j \in A}$ using Neumark's theorem,
we would like to implement the unitary transformation $U^x$ with the
upper left submatrix
\be
  V^x = \frac{1}{\sqrt{|A|}} \sum_{j,w \in A} \carbar{w}{j} \, |j\>\<S^x_w|
\,.
\ee
It is convenient to perform a Fourier transform (over $A$) on the
left (i.e., on the index $j$), giving a unitary operator $\tilde U^x$
with upper left submatrix
\be
  \tilde V^x = \frac{1}{|A|} \sum_{j,w,v \in A} 
               \car{w}{j} \, \carbar{v}{j} \, |w\>\<S^x_v|
             = \sum_{w \in A} |w\>\<S^x_w|
\,.
\ee
Therefore, the \PGM can be implemented efficiently if we can
efficiently perform the transformation
\be
  |x,w\> \mapsto \begin{cases} |x,S^x_w\> & \eta^x_w > 0 \\
                               |\xi^x_w\> & \eta^x_w = 0 \end{cases}
\label{eq:qsample}
\ee
where $|\xi^x_w\>$ is any state allowed by the unitarity of $\tilde
U^x$.  We refer to (\ref{eq:qsample}) as \emph{quantum sampling} of
solutions to the matrix sum problem.  If we can efficiently quantum
sample from matrix sum solutions, then by running the circuit in
reverse, we can efficiently implement $\tilde U^x$, and hence the
desired measurement.

\section{Metacyclic groups}
\label{sec:metacyclic}

In this section, we present our first application of the \PGM approach
to a particular class of groups, those of the form $\ZN \semidirect
\Zp$ with $p$ prime.  All such groups are {\em metacyclic}.  The
possible automorphisms $\func$ correspond to multiplication by some
$\mat \in \ZN^\times$, so that $\func(a)=\mat a$.  For $\func^p$ to be
the identity map, we must have $\mat^p = 1 \bmod N$.  The function
$\fsum{b}$ can be represented by the sum
\be
  \msum{b} := \sum_{i=0}^{b-1} \mat^i
\,,
\label{eq:geom}
\ee
so that $\fsum{b}(a)=\fsumt{b}(a)=\msum{b}a$.

Note that because $\<\mat\> \le \ZN^\times$, Lagrange's theorem
implies that $p$ divides $|\ZN^\times|=\phi(N)$, where $\phi(N)$
denotes the Euler totient function of $N$, i.e., the number of
elements of $\ZN$ that are relatively prime to $N$.  To give an
efficient algorithm, we require that $N$ and $p$ be fairly close; in
particular, we require $N/p = \poly(\log N)$.

\subsection{Solution of the matrix sum problem}

To apply the \PGM to the \HSP over groups of the form $\ZN \semidirect
\Zp$, we must understand the matrix sum problem for such groups.  In
general, the problem is the following: given uniformly random $x \in
\ZN^k$ and $w \in \ZN$, find $b \in \Zp^k$ such that $\sum_{j=1}^k
\msum{b_j} x_j = w$.  We note in passing that if $\mat-1 \in
\ZN^\times$ (which is always the case if, for example, $N$ is prime),
then we can sum the geometric series (\ref{eq:geom}) to obtain
$\msum{b} = (\mat^b - 1)/(\mat-1)$,
and in particular, $\msum{p} = 0$, so that $|\<(d,1)\>|=p$ for all $d
\in \ZN$.  However, for the following, we do not need to assume that
$N$ is prime, that $\mat-1 \in \ZN^\times$, or even that $\msum{p}=0$.

For $k=1$, the matrix sum problem $\msum{b} x = w$ is particularly
easy to solve.  Assume that $x \in \ZN^\times$, which occurs with
probability $\phi(N)/N \in \Omega(1/\log\log N)$, so that we can
rewrite the equation as $\msum{b} = w/x$.

If such a $b$ exists, it must be unique, since $\msum{b} \ne
\msum{b'}$ for $b \ne b' \in \Zp$, which can be seen as follows.
Supposing the claim is false, with $b>b'$ without loss of generality,
$\msum{b}-\msum{b'} = \mat^{b'} \msum{b-b'} = 0$, so $\msum{b-b'}=0$.
For $\mu \ne 1$, $\<\mu\> \le \ZN^\times$ has $|\<\mu\>|=p$ since
$\mu^p = 1$.  Using the identity
\be
  (\mat-1)\msum{b} = \mat^b - 1
\,,
\label{eq:geomsumtrick}
\ee
it is clear that $\msum{b-b'}=0$ implies $\mat^{b-b'} = 1$, so that
$b-b'$ is a multiple of $p$.  But since $b,b' \in \Zp$ with $b \ne
b'$, this is a contradiction.

Now from (\ref{eq:geomsumtrick}), we see that by solving the discrete
logarithm problem $\mat^b = 1+(\mat-1)w/x$ for $b$, we solve the
matrix sum problem.  The discrete logarithm can be calculated
efficiently using Shor's algorithm \cite{Shor:94a}, which implies that
the unique solution to the matrix sum problem can be found efficiently
when it exists.

For the \PGM to be successful with reasonably high probability, the
equation $\msum{b} = w/x$ must be likely to have a solution.  Since
the various values of $b \in \Zp$ yield $p$ distinct possibilities for
$\msum{b}$, and since $w/x$ is uniformly random in $\ZN$ under the
assumption $x \in \ZN^\times$, a solution exists with probability
$p/N$, which is not too small given $N/p = \poly(\log N)$.  Taking
into account the probability that $x \in \ZN^\times$, a solution
exists with probability at least $\phi(N)p/N^2$.  Therefore, by
Lemma~\ref{lem:sucprob}, the \PGM succeeds with probability
$1/\poly(\log Np)$, and we have

\begin{theorem}
\label{thm:metacyclic}
For $p$ prime and $N$ arbitrary with $N/p=\poly(\log N)$,
the hidden subgroup problem over $\ZN \semidirect \Zp$ can be
solved in time $\poly(\log Np)$.
\end{theorem}
\noindent
This generalizes a result of \cite{Moore:04a}, which proves
Theorem~\ref{thm:metacyclic} in the case where $N$ is prime.

If $N/p$ were superpolynomial, then we could imagine solving the
problem by implementing the \PGM with $k>1$ copies of the hidden
subgroup state.  However, in this case, the matrix sum problem is not
simply a discrete logarithm, so it is not clear whether it can be
solved efficiently.

\subsection{Stripped down algorithm}

Since the \PGM approach described above requires only $k=1$ copy of
the hidden subgroup state, it yields a very simple quantum algorithm.
Here we present this result without reference to the general
framework, giving an algorithm that is quite straightforward,
especially in comparison to the earlier algorithm of \cite{Moore:04a},
which requires a nonabelian Fourier transform.

Consider the order $p$ hidden subgroup $H = \{(\msum{b} d,b) :
b\in\Zp\}$ for $d \in \ZN$ with $\msum{p}d=0$.  The left cosets are of
the form $(\ell,0)H$ for $\ell \in \ZN$, and the corresponding coset
states (\ref{eq:cosetstates}) are
\be
  |\psi_{\ell,d}\> 
    = \frac{1}{\sqrt{p}}\sum_{b\in\Zp} |\ell + \msum{b} d,b\>
\,.
\ee
Applying the Fourier transform over $\ZN$ to the first register,
we obtain
\be
  |\tilde{\psi}_{\ell,d}\> 
    = \frac{1}{\sqrt{Np}} \sum_{x\in\ZN} \sum_{b\in\Zp}
      \omega^{x(\ell + \msum{b} d)} |x,b\>
\ee
where $\omega := \exp{(2\pi \i/N)}$.  Now measure the first register
and assume that the result is some $x \in \ZN^\times$, which happens
with probability $\phi(N)/N$.  Then append an ancilla register to this
state and perform the calculation $|b,0\> \mapsto |b,x \msum{b}\>$,
giving the state
\be
  \frac{1}{\sqrt p} \sum_{b\in\Zp} \omega^{x \msum{b} d} |b,x\msum{b}\>
\,.
\ee
Note that for any fixed $b \in \Zp$, $\msum{b}$ can be calculated
efficiently by repeated squaring, using the fact that $\msum{2b} =
(1+\mat^b)\msum{b}$.

Now we perform a classical computation to erase the value of $b$.
With the expression (\ref{eq:geomsumtrick}) and knowledge of $x \in
\ZN^\times$ and $\mat$, we can efficiently perform the computation
$|x \msum{b},0\> \mapsto |x \msum{b},\mat^b\>$.  Then, using Shor's
discrete logarithm algorithm, we efficiently compute $|x
\msum{b},\mat^b\> \mapsto |x\msum{b},b\>$.  Using this procedure to
erase the value of $b$, we see that we can produce the state
\be
  \frac{1}{\sqrt p} \sum_{b\in\Zp}
  \omega^{x \msum{b} d} |x \msum{b} \>
\,.
\label{eq:actual}
\ee

Finally, we perform an inverse Fourier transform over $\ZN$ and
observe the register in the hope of obtaining $d$.  To calculate the
probability of this happening, consider the perfect state
\be
  |\tilde d\> := \frac{1}{\sqrt N} \sum_{j\in\ZN}\omega^{jd} |j\>
\,,
\label{eq:perfect}
\ee
which would be guaranteed to yield the answer $d$.  The overlap
between the perfect state (\ref{eq:perfect}) and the actual state
(\ref{eq:actual}) is $\sqrt{p/N}$, so the probability of observing $d$
is at least $p/N$.  Thus, overall, we find a success probability of at
least $\phi(N) p/N^2$.  Because $\phi(N)/N\in \Omega(1/\log\log N)$,
and $N/p = \poly(\log N)$ by assumption, simply repeating the above
protocol $N^2/\phi(N)p = \poly(\log Np)$ times gives an efficient
quantum algorithm for the \HSP over $\ZN \semidirect \Zp$.

\section{Groups of the form \boldmath$\Zp^r \semidirect \Zp$}
\label{sec:heisesque}

We now turn to another class of semidirect product groups, those of
the form $\Zp^r \semidirect_\varphi \Zp$.  We find an efficient
quantum algorithm for such groups for any $\func$, so long as $r$ is
constant.  In other words, the running time of the algorithm is
$\poly(\log p)$.  The algorithm is particularly simple in the case
$r=2$, where the only nontrivial semidirect product is known as the
Heisenberg group (for which it was recently shown that there is an
efficient quantum algorithm whose output information theoretically
determines the solution of the \HSP \cite{RRS:05}).  We present the
algorithm for $r=2$ in Section \ref{sec:heisenberg}, and then proceed
to the general case in Section \ref{sec:heisgen}.

In general, when $A$ is the elementary abelian $p$-group $A = \Zp^r$,
its automorphism group is $\Aut(\Zp^r) \cong \GL_r(\Zp)$.  Therefore,
$\func$ can be identified with a nonsingular matrix $\mat \in
\GL_r(\Zp)$ such that $\mat^p = I$.  As before, we define $\msum{b} :=
\sum_{i=0}^{b-1} \mat^i$ so that $\fsumt{b}(a)=\msum{b}a$ (see
Appendix~\ref{app:matrep}).

\subsection{The Heisenberg group}
\label{sec:heisenberg}

For $r=2$, there are only two nonisomorphic semidirect product groups
$\Zp^2 \semidirect \Zp$: the abelian group $\Zp^3$ and the {\em
Heisenberg group}, for which
\be
  \mat = \begin{pmatrix}1 & 1 \\ 0 & 1\end{pmatrix}
\,.
\ee
The matrix sum problem for the Heisenberg group is the following:
given uniformly random $x,y \in \Zp^k$ and $w,v \in \Zp$, find $b \in
\Zp^k$ such that
\be
  \sum_{j=1}^k \sum_{i=0}^{b_j-1}{\mat^i} \begin{pmatrix}x_j \\ y_j
  \end{pmatrix}
  = \begin{pmatrix}w \\ v\end{pmatrix}
\,.
\ee
Clearly,
\be
  \mat^i = \begin{pmatrix}1 & i \\ 0 & 1\end{pmatrix}
  \quad \text{so}\quad
  \msum{b} := \sum_{i=0}^{b-1} \mat^i =
  \begin{pmatrix} b & sb(1-b) \\ 0 & b \end{pmatrix}
\ee
where $s$ is the multiplicative inverse of $-2$ in the finite field
$\Zp$, defined by $p=2s+1$.  Therefore the matrix sum problem can be
written
\ba
  \sum_{j=1}^k \begin{pmatrix} b_j & sb_j(1-b_j) \\ 0 & b_j \end{pmatrix}
  \begin{pmatrix}x_j \\ y_j\end{pmatrix}
  = \begin{pmatrix}w \\ v\end{pmatrix}
\,.
\ea
If $k=1$, then it is not hard to see that the probability of having a
solution is $O(1/p)$, i.e., exponentially small in $\log p$.  However,
if we take $k=2$, then there are as many variables as (scalar)
equations, and we find the matrix sum problem
\ba
  b_1 x_1 + s b_1(1-b_1)y_1 + b_2 x_2 + s b_2(1-b_2)y_2 &= w
  \label{eq:heismsp1}\\
                                      b_1 y_1 + b_2 y_2 &= v
  \label{eq:heismsp2}
\ea
with uniformly random $x_1,x_2,y_1,y_2,w,v \in \Zp$.

The equations (\ref{eq:heismsp1}--\ref{eq:heismsp2}) can be solved as follows.  Define
\be
  \Delta := (2wy_1+vy_1-v^2-2vx_1)(y_1+y_2)y_2+(v y_2 +x_1 y_2-x_2y_1)^2
\,.
\ee
If $\Delta$ is a nonzero square in $\Zp$ and $y_1,y_2,y_1+y_2 \ne 0$,
then there are two solutions for $(b_1,b_2)$:
\be
  b_1 = \frac{v y_1+x_2y_1-x_1y_2\pm\sqrt{\Delta}}{y_1(y_1+y_2)}
  \quad \text{and} \quad
  b_2 = \frac{v y_2+x_1y_2-x_2y_1\mp\sqrt{\Delta}}{y_2(y_1+y_2)}
\,.
\ee
It is straightforward to check that $(b_1,b_2)$ is indeed a solution.
If $\Delta=0$ then these two solutions are the same, and there is only
one solution.  Finally, if $\Delta$ is nonsquare, then there are no
solutions.

To calculate the success probability of the \PGM, we must determine the
probability that $y_1,y_2,y_1+y_2 \ne 0$ and $\Delta$ is a square in
$\Zp$.  First, note that since $y_1,y_2$ are uniformly random, we
have $y_1,y_2,(y_1+y_2)\neq 0$ with probability $(p-1)(p-2)/p^2 =
1-O(1/p)$.  Now rewrite $\Delta$ as
\be
  \Delta = 2wy_1y_2(y_1+y_2) +\left[v(y_1-2x_1-v)(y_1+y_2)y_2
           +(vy_2+x_1 y_2-x_2y_1)^2\right]
\,.
\ee
Assuming $2y_1y_2(y_1+y_2)\neq 0$, $\Delta$ depends linearly on $w$.
Hence, if we fix $x_1,x_2,y_1,y_2,v$ and choose $w$ uniformly at
random from $\Zp$, then $\Delta$ will also be uniformly random in
$\Zp$.  Of the possible values of $\Delta \in \Zp$, we have $(p-1)/2$
cases with $\sqrt\Delta \ne 0$, $(p-1)/2$ cases with $\Delta$ not a
square, and one case with $\Delta=0$.  Therefore, under the assumption
$y_1,y_2,y_1+y_2 \ne 0$, we have
\ba
  \Pr(\eta^{x,y}_{w,v} = 0) &= \textstyle\frac{1}{2}-\frac{1}{2p} \\
  \Pr(\eta^{x,y}_{w,v} = 1) &= \textstyle\frac{1}{p} \\
  \Pr(\eta^{x,y}_{w,v} = 2) &= \textstyle\frac{1}{2}-\frac{1}{2p}
\,.
\ea
In particular, we see that $\Pr(\eta^{x,y}_{w,v} = 2) =
\frac{1}{2}-O(1/p)$.  Therefore, by Lemma~\ref{lem:sucprob}, $\Prsuc
\ge 1-O(1/p)$.

Since the above discussion gives an explicit solution of the matrix
sum problem, and since arithmetic in the finite field $\Zp$ can be
performed in time $\poly(\log p)$, it is straightforward to
efficiently implement the quantum sampling transformation
(\ref{eq:qsample}).\footnote{For example, given the ability to compute
a list of solutions $b_1,\ldots,b_\eta$ with $\eta$ small, the
following simple trick can be used to efficiently create the uniform
superposition.  Create the labeled superposition $\sum_{j \in
\Z_\eta} |j,b_j\>/\sqrt\eta$, Fourier transform the first register
(over $\Z_\eta$), and measure the first register.  When the outcome is
$0$, which occurs with probability $1/\eta$, the desired state is
obtained.  Finally, use $O(\eta)$ repetitions to boost the success
probability close to $1$, and implement the measurement unitarily.}
Combined with the fact that the success probability of the \PGM is
large, this shows that the hidden subgroup problem in the Heisenberg
group can be solved efficiently.

\subsection{The general case}
\label{sec:heisgen}

More generally, consider the group $\Zp^r \semidirect_\varphi \Zp$ for
any constant $r$ and for any automorphism $\func$, i.e., for any
matrix $\mat$ satisfying $\mat^p=I$.  It is easy to see that matrices
related by a similarity transformation correspond to isomorphic
groups.  Thus, without loss of generality, we can assume that $\mat$
is in Jordan canonical form, or in other words, that it is zero
everywhere except the diagonal and first superdiagonal, and that the
elements on the first superdiagonal are either $0$ or $1$.  Since
$\mat^p = I$, the diagonal elements of this $\mat$ must all be equal
to $1$.  Thus the various nonisomorphic groups of the form $\Zp^r
\semidirect \Zp$ correspond to the partitions of $r$, where the
partitions describe the sizes of the Jordan blocks.

For simplicity, we consider the case of a single Jordan block of size
$r$; the extension to other cases will be clear.  In other words, we
consider the $r \times r$ matrix
\be
  \mat = \begin{pmatrix}
    1 & 1 & 0 & \cdots & 0 \\
    0 & 1 & 1 & \ddots & \vdots \\
    0 & 0 & \ddots & \ddots & 0 \\
    \vdots & \ddots & \ddots & 1 & 1 \\
    0 & \cdots & 0 & 0 & 1 \vdotsstrut
  \end{pmatrix}
\,.
\ee
Then the $b$th matrix sum $\msum{b} :=
I+\mat+\mat^2+\cdots+\mat^{b-1}$ is given by
\be
  \msum{b} = \begin{pmatrix}
           \binom{b}{1} & \binom{b}{2} & \binom{b}{3} & \cdots & \binom{b}{r} \\
           0 & \binom{b}{1} & \binom{b}{2} & \ddots & \vdots \\
           0 & 0 & \ddots & \ddots & \binom{b}{3} \\
           \vdots & \ddots & \ddots & \binom{b}{1} & \binom{b}{2} \\
           0 & \cdots & 0 & 0 & \binom{b}{1} \vdotsstrut
         \end{pmatrix} \bmod p \,.
\label{eq:matsum}
\ee
Note that $\msum{0}=\msum{p}=0$.  If $\mat$ consists of several Jordan
blocks, then (\ref{eq:matsum}) can be applied in each block.

The matrix sum problem is thus the following: given uniformly random
$x \in (\Zp^r)^k$ and $w \in \Zp^r$, find $b \in \Zp^k$ such that
$\sum_{j=1}^k \msum{b_j} x_j=w$.  This is a set of $r$ polynomial
equations over $\Zp$ in $k$ variables.  For example, in the case of a
single Jordan block, we have
\be
  \sum_{j=1}^k \bigg[\binom{b_j}{1} x_{i,j}+\cdots+\binom{b_j}{i} x_{1,j}\bigg]
    = w_i
    \label{eq:msumeqi}
\ee
for all $1\leq i \leq r$, where the $i$th equation is of degree $i$.

If $k<r$, so that there are fewer unknowns than equations, we expect
that the matrix sum problem will typically have no solutions.  On the
other hand, if $k>r$, we expect that the ideal generated by
(\ref{eq:msumeqi}) will typically have dimension
$k-r$, and there will be $O(p^{k-r})$ solutions.  By choosing $k=r$,
we can ensure that there are typically $O(1)$ solutions, so that the
\PGM succeeds, yet the solutions of the matrix sum problem are few in
number and thus relatively easy to find (and to quantum sample).

This intuition can be formalized by calculating the mean and variance
of the number of solutions.  Using such an argument, we find the
following:
\begin{lemma}
For the matrix sum problem of any group $\Zp^r \semidirect \Zp$ with
$k=r$, $\Pr(\eta^x_w=1 \text{~or~} 2) \ge \frac{1}{4}$.
\label{lem:heisesqueprob}
\end{lemma}

\begin{proof}
See Appendix~\ref{app:heisesqueprob}.
\end{proof}

By Lemma~\ref{lem:sucprob}, this shows that $\Prsuc \ge 1/16$, even if
we only consider the cases in which the number of solutions is at most
$2$.  Thus, the pretty good measurement succeeds in identifying the
hidden subgroup.

To efficiently implement the \PGM, we quantum sample from solutions of
the matrix sum problem, which can be done by computing a list of all
solutions.  A list of solutions can be found by first using
Buchberger's algorithm to compute a Gr\"obner basis for the ideal and
then using elimination theory \cite{CLO:97}.  In general, upper
bounding the complexity of Buchberger's algorithm is a difficult
problem, and it is known that the algorithm may be very inefficient in
terms of the number of variables, the number of equations, and the
degree of those equations.  However, since we consider $r$ constant,
such inefficiency is not an issue.  The running time in terms of $p$,
the size of the field, enters only as an overall $\poly(\log p)$
factor, accounting for the cost of performing arithmetic operations in
a finite field (see for example \cite{DMY:95}).  Thus, for our
purposes, a Gr\"obner basis can be computed efficiently.  Since we
consider only the cases in which there are at most $2$ solutions,
elimination is also efficient, giving an overall $\poly(\log p)$ time
algorithm for computing a list of matrix sum solutions, and hence for
quantum sampling from those solutions.

Collecting these results, we find
\begin{theorem}
The hidden subgroup problem over any group of the form $\Zp^r
\semidirect \Zp$ with $r$ fixed can be solved in time $\poly(\log p)$
on a quantum computer.
\end{theorem}

\section{Discussion}
\label{sec:discussion}

In this paper, we have studied the pretty good measurement for
semidirect product groups of the form $A \semidirect \Zp$ with $A$
abelian and $p$ prime.  We found that the \PGM is closely connected to
the matrix sum problem, and we exploited this connection to find
efficient quantum algorithms for certain metacyclic groups (Section
\ref{sec:metacyclic}) as well as all groups of the form $\Zp^r
\semidirect \Zp$ with $r$ fixed (Section \ref{sec:heisesque}).  The
latter algorithm demonstrates that entangled measurements may be
useful for efficiently solving the nonabelian \HSP.

Aside from the fact that these particular nonabelian groups admit
efficient quantum algorithms, our results suggest two general
directions for further investigation.  First, in the standard approach
to the nonabelian \HSP, it would be helpful to have a better
understanding of when entangled measurements are necessary and when
they can be implemented to give efficient algorithms.  Second, for the
\HSP or for other problems that can be viewed as quantum state
distinguishability problems, identifying an optimal measurement (or
considering a particularly nice measurement such as the \PGM, which in
general may or may not be optimal) can be used as a principle for
discovering new quantum algorithms.

While the reduction of Lemma~\ref{lem:reduction} combined with the
\PGM approach outlined in Section~\ref{sec:pgm} appears to efficiently
solve the \HSP in most of the semidirect product groups where
efficient algorithms are known, there is one notable exception.
The groups $\Zp^n \semidirect \Z_2$ (for which an efficient quantum
algorithm is given in \cite{Friedl:03a}) give rise to a subset sum
problem over $\Zp^n$, which appears to be essentially as difficult as
the subset sum problem over $\ZN$ arising from the dihedral group.
Thus, it would be interesting to understand what allows the algorithm
in \cite{Friedl:03a} to be efficient even though the matrix sum
problem is (apparently) hard.

Of course, there are many nonabelian groups that are semidirect
products of nonabelian groups, or that cannot be nontrivially
decomposed into semidirect products at all.  The \PGM approach is well
defined for any group, so it would be interesting to explore the
approach in such cases, regardless of whether the \PGM is optimal.

\paragraph*{Acknowledgments}

We thank Carlos Mochon and Frank Verstrate for helpful discussions of
Theorem~\ref{thm:holevo}.
AMC received support from the National Science Foundation under Grant
No.\ EIA-0086038.


\appendix

\section{Matrix representation of \boldmath$\fsum{b}$}
\label{app:matrep}

In this appendix, we show that $\fsum{b}$ and $\fsumt{b}$ can be
represented by (simply related) matrices, and furthermore, that
$\car{x}{\fsum{b}(d)} = \car{\fsumt{b}(x)}{d}$.

\begin{lemma}
\label{lemma:conjPhi}
Let $A$ be a finite abelian group, $\varphi\in\Aut(A)$, $b\in\N$ with
the corresponding
$\Phi:A \to A$ defined by $\Phi := \sum_{i=0}^{b-1}{\varphi^i}$,
and $\chi:A\rightarrow\C$
a character of $A$.  Then there exists a function
$\hat\Phi:A\rightarrow A$ such that
$\car{x}{\Phi(d)} = \car{\hat\Phi(x)}{d}$ for all $d,x\in A$.
\end{lemma}

\begin{proof}
Let $A \cong A_{p_1} \times \cdots \times A_{p_r}$ be the elementary
divisor decomposition of $A$, i.e., the decomposition into $p$-groups
where the $p_i$ are distinct primes.  Accordingly, let
$d=(d_1,\dots,d_r)$, $x=(x_1,\dots,x_r)$, and $\chi^{(i)}_{x_i}:
A_{p_i} \to \C$ such that $\car{x}{d} =
\chi^{(1)}_{x_1}(d_1)\cdots\chi^{(r)}_{x_r}(d_r)$.  It is known that
any automorphism $\varphi \in \Aut(A)$ can be decomposed as $\varphi
=(\varphi_1,\dots,\varphi_r)$ with $\varphi_i \in \Aut(A_{p_i})$
\cite{Shoda}.  Similarly, we have the decomposition $\Phi(d) =
(\Phi_1(d_1),\dots,\Phi_r(d_r))$.  Hence if we prove the lemma for
$p$-groups, then this proves it for all finite abelian groups.

Assume therefore that $A$ is a $p$-group, $A \cong \Z_{p^{e_1}} \times
\cdots \times \Z_{p^{e_k}}$ with $e_1 \ge \cdots \ge e_k$.  By
\cite{Shoda}, we can represent $\varphi \in \Aut(A)$ as a matrix
transformation $(x_1,\dots,x_k) \mapsto (x_1,\dots,x_k)\mu$ where $\mu
\in \Z^{k\times k}$ and $p^{e_j-e_i}|\mu_{ij}$ for all $i>j$.
Consequently, we can also represent the transformation $\Phi$ by a
matrix $M$ with $p^{e_j-e_i}|M_{ij}$ for all $i>j$.  Now define a
conjugate matrix $\hat M_{ji} := p^{e_i-e_j} M_{ij}$ for all $i,j$
(note that all entries of $\hat M$ are integers).  For any character
$\chi_x: A \to \C$, we find
\be
  \chi_x(\Phi(d)) =
  \exp\Big(2\pi\i\sum_{ji}{d_i M_{ij} x_j/p^{e_j}}\Big) =
  \exp\Big(2\pi\i\sum_{ij}{d_i x_j \hat M_{ji}/p^{e_i}}\Big) =
  \chi_{\hat\Phi(x)}(d)
\,,
\ee
where $\hat\Phi: A \to A$ is the matrix transformation defined by
$(x_1,\dots,x_k) \mapsto (x_1,\dots,x_k)\hat M$.
\end{proof}

Note that if $A=\Z_N$ (as in Section~\ref{sec:metacyclic}), then
$\Phi$ can be represented by a single scalar $M \in \Z_N$, and hence
$\hat\Phi=\Phi$.  Also, if $A = \Zp^r$ (as in
Section~\ref{sec:heisesque}), then $M \in \Zp^{r \times r}$ and $\hat
M = M^T$.  For notational simplicity, since we only need the matrix
representation of $\hat\Phi$ (and not of $\Phi$) in these two cases,
we put $\mat \to \mat^T$ and $M \to M^T$ throughout the body of the
paper.

\section{Proof of Lemma~\ref{lem:heisesqueprob}}
\label{app:heisesqueprob}

\begin{proof}
For $x \in (\Zp^n)^k$, $w \in \Zp^n$, and $b \in \Zp^k$, let
$\Xi^x_w(b)=0$ denote the system of polynomial equations
(\ref{eq:msumeqi}).  We want to understand the typical behavior of
\be
  \eta^x_w := |\{b: \Xi^x_w(b)=0\}| = \sum_b \delta[\Xi^x_w(b)=0]
\ee
for uniformly random $x,w$.  Specifically, we want to show that it is
typically close to its mean,
\be
  \mu := \E_{{x \in A^k, w \in A}}[\eta^x_w]
      = p^{k-r}
\ee
where we have used (\ref{eq:exeta}).  To compute the variance, note
that
\ba
  \E_{{x \in A^k, w \in A}}[(\eta^x_w)^2]
    &= \frac{1}{p^{rk+r}} \sum_{x,w} (\eta^x_w)^2 \\
    &= \frac{1}{p^{rk+r}} \sum_{x,w}
       \bigg(\sum_b \delta[\Xi^x_w(b)=0]\bigg)
       \bigg(\sum_c \delta[\Xi^x_w(c)=0]\bigg) \\
    &= \frac{1}{p^{rk+r}} \sum_{x,w}
       \bigg( \sum_b \delta[\Xi^x_w(b)=0]
             +\sum_{b \ne c} \delta[\Xi^x_w(b)=\Xi^x_w(c)=0] \bigg)
\,.
\ea
The first (diagonal) term is just the mean.  For the second
(off-diagonal) term, note that the condition $\Xi^x_w(b)=\Xi^x_w(c)$
actually does not depend on $w$, so we can write
\ba
  \E_{{x \in A^k, w \in A}}[(\eta^x_w)^2]
    &= \mu + \frac{1}{p^{rk+r}} \sum_{b \ne c} \sum_x
       \delta[\Xi^x_0(b)=\Xi^x_0(c)] \sum_w \delta[\Xi^x_w(b)=0] \\
    &= \mu + \frac{1}{p^{rk+r}} \sum_{b \ne c} \sum_x
       \delta[\Xi^x_0(b)=\Xi^x_0(c)]
\ea
where we have used the fact that for any fixed $x,b$, there is exactly
one $w$ that satisfies the equation.  Now for $b \ne c$, choose some
$j$ such that $b_j \ne c_j$.  The system of equations
$\Xi^x_0(b)=\Xi^x_0(c)$ is linear in $x$, and in the $i$th equation
(\ref{eq:msumeqi}), the coefficient of $x_{ij}$ is nonzero (it is
simply $b_j - c_j$), so for any values of the $x_{ij'}$ for $j' \ne
j$, we can solve the equations uniquely for the $x_{ij}$, giving
\ba
  \E_{{x \in A^k, w \in A}}[(\eta^x_w)^2]
    &= \mu + \frac{1}{p^{rk+r}} \sum_{b \ne c} \sum_x
       \delta[\forall i ~ x_{ij} \text{~fixed}] \\
    &= \mu + \frac{1}{p^{rk+r}} \sum_{b \ne c} p^{rk-r} \\
    &= \mu + p^{2(k-r)} - p^{k-2r}
\,.
\ea
Thus we find
$  \sigma^2 := \E_{x,w}[(\eta^x_w)^2] - \mu^2
           =  \mu - p^{k-2r}
           =  \mu (1-p^{-r})$.

Since the variance is small, Chebyshev's inequality shows that the
probability of deviating far from the mean number of solutions is
small:
\be
  \Pr(|\eta^x_w - \mu| \ge c) \le \frac{\sigma^2}{c^2}
\,.
\label{eq:cheby}
\ee
For $k=r$, we have
$\mu = 1$ and $\sigma^2 = 1 - p^{-r}$,
so by putting $c=2$ in (\ref{eq:cheby}), we find
$\Pr(\eta^x_w \ge 3) \le \frac{1}{4}$.

To see that we are unlikely to have no solutions, we need a slightly
stronger bound than the Chebyshev inequality.  Since $\eta^x_w$ is a
nonnegative, integer-valued random variable, we
have \cite[p.~58]{AS:00}
$\Pr(\eta^x_w = 0) \le {\sigma^2}/({\mu^2 + \sigma^2})
=   (1 - p^{-r})/(2-p^{-r}) \le 1/2$.
Combining these results, we see that
$\Pr(\eta^x_w = 1 \text{~or~} 2) \ge \frac{1}{4}$
as claimed.
\end{proof}

\end{document}